\documentclass{article}

\usepackage{PRIMEarxiv}

\usepackage[utf8]{inputenc} 
\usepackage[T1]{fontenc}    
\usepackage{hyperref}       
\usepackage{url}            
\usepackage{booktabs}       
\usepackage{amsfonts}       
\usepackage{nicefrac}       
\usepackage{microtype}      
\usepackage{lipsum}
\usepackage{threeparttable}
\usepackage{fancyhdr}       
\usepackage{graphicx}       
\graphicspath{{media/}}     

\title{AI in Pakistani Schools: Adoption, Usage, and Perceived Impact among Educators}

\author{
  Syed Hasan Raza \\
  Miami University \\
  Oxford, OH \\
  \texttt{razash@miamioh.edu} \\
   \And
  Azib Farooq \\
  Miami University \\
  Oxford, OH \\
  \texttt{farooqa@miamioh.edu} \\
}

\begin{document}
\maketitle

\begin{abstract}
  Artificial Intelligence (AI) is increasingly permeating classrooms worldwide, yet its adoption in schools of developing countries remains under-explored. This paper investigates AI adoption, usage patterns, and perceived impact in Pakistani K–12 schools based on a survey of 125 educators. The questionnaire covered educator's familiarity with AI, frequency and modes of use, and attitudes toward AI’s benefits and challenges. Results reveal a generally positive disposition towards AI: over two-thirds of teachers expressed willingness to adopt AI tools given proper support and many have begun integrating AI for lesson planning and content creation. However, AI usage is uneven – while about one-third of respondents actively use AI tools frequently, others remain occasional users. Content generation emerged as the most common AI application, whereas AI-driven grading and feedback are rarely used. Teachers reported moderate improvements in student engagement and efficiency due to AI, but also voiced concerns about equitable access. These findings highlight both the enthusiasm for AI’s potential in Pakistan’s schools and the need for training and infrastructure to ensure inclusive and effective implementation.
\end{abstract}


\maketitle

\renewcommand\thefootnote{}

\renewcommand\thefootnote{\fnsymbol{footnote}}
\setcounter{footnote}{1}

\section{Introduction}\label{sec:introduction}

AI technologies are rapidly transforming education on a global scale, promising personalized learning and automated support for teachers and students \cite{harry2023role}. Countries like China and India have extensively implemented AI in education, using intelligent tutoring systems, content recommendation, and even humanoid robots, which has enhanced student engagement and learning outcomes \cite{niu2022teachers, kasinathan2019exploring, shoukat2024harnessing}. In many developing regions, high student-to-teacher ratios and shortages of qualified teachers drive interest in AI as a means to bridge educational gaps. For instance, AI-powered tutoring and personalized learning tools are seen as necessary innovations to improve equity in lower-income school systems \cite{Fan_2024_TippingTheScales}. Reflecting the global trend, the adoption of generative AI tools in classrooms has accelerated dramatically – reaching millions of users within months – signaling educator's eagerness to experiment with new technologies \cite{Fan_2024_TippingTheScales, kim2025perceptions}. This surge in AI interest has set the stage for a potential paradigm shift in teaching and learning practices worldwide.

At the same time, the integration of AI in K–12 education is still in its early stages, especially in developing countries. Educators and institutions are grappling with varying levels of AI familiarity, preparedness, and ethical considerations as they consider these emerging tools \cite{kim2025perceptions}. Pakistan’s education sector exemplifies this dual reality of promise and challenge. National policymakers have acknowledged AI’s potential to improve educational outcomes, and initial pilot projects (such as AI-driven tutoring platforms) have demonstrated positive impacts on student learning \cite{shoukat2024harnessing}. Yet significant barriers persist: many Pakistani schools face limited access to AI technologies due to cost constraints and inadequate infrastructure \cite{shoukat2024harnessing}. Furthermore, without proper training and support, teachers may struggle to integrate AI effectively or ethically. Recent studies emphasize that issues like insufficient professional development, data privacy concerns, and infrastructural inequities can hinder teachers’ adoption of AI \cite{kim2025perceptions}. Thus, understanding how Pakistani educators are beginning to adopt and use AI – and what challenges they face – is crucial for guiding policy and support. 

This study addresses that need by examining the current state of AI adoption, use, and perceived impact in Pakistani schools, offering insights into how teachers in a developing country context are engaging with AI in their classrooms.
\section{Literature Review} \label{sec:literature}

Research on AI in education highlights a mix of optimism and caution. Proponents argue that AI has the potential to revolutionize classroom instruction by enabling personalized learning at scale, automating routine tasks, and providing intelligent tutoring support \cite{holmes2022state, zhai2021review}. Empirical evidence from various contexts suggests that AI tools can enhance teaching efficiency and student outcomes. For example, McKinsey estimates that existing AI technologies could allow teachers to reallocate 20–40\% of their time to higher-value activities by taking over administrative and grading tasks \cite{Fan_2024_TippingTheScales}
. In practice, many educators have started leveraging AI for content creation and lesson planning, finding that tools like chatbots and content generators help diversify instructional materials and address diverse learning needs \cite{ahmed2024embracing}. A recent meta-analysis of teacher technology acceptance found that perceived usefulness (e.g. improvements in student learning) and ease of use significantly predict educator's intentions to adopt AI \cite{mulyani2025transforming}. This aligns with emerging reports noted that ChatGPT was used by all respondents, with daily AI use common among those teachers \cite{ahmed2024embracing}. Such findings illustrate a growing enthusiasm in the teaching community to experiment with AI when it clearly adds value in the classroom.

On the other hand, the literature also underscores several barriers and concerns that accompany AI’s entrance into schools. Educator preparedness is a recurring theme – teachers often feel they lack the training and confidence to integrate AI tools into pedagogy effectively \cite{kim2025perceptions}. Without formal professional development, many resort to self-learning or avoid using AI beyond basic trials. Another concern involves ethical and equity issues: AI systems can inadvertently carry biases or widen disparities if not carefully managed. For instance, access to AI-powered resources may be uneven, exacerbating the digital divide between well-resourced and under-resourced schools \cite{ahmed2024digital}. UNESCO and other bodies have stressed that AI in education must be guided by policies ensuring equitable access and data privacy \cite{shoukat2024harnessing}. In Pakistan, recent policy discussions acknowledge that while AI could significantly improve education, its integration must overcome hurdles of infrastructure and cost \cite{shoukat2024harnessing}. Studies from similar contexts (e.g. African and South Asian schools) have reported that unreliable internet connectivity and lack of devices impede teacher's ability to utilize AI tools, even when they are willing to try them \cite{kim2025perceptions}. Additionally, teachers express concern about maintaining a human touch in teaching – they seek to balance AI assistance with personal interaction and judgement. 

In summary, prior research suggests that successful AI adoption in schools requires not only the availability of effective AI tools, but also supportive training, robust infrastructure, and clear guidelines to address ethical considerations. These insights from the literature form the backdrop for our investigation into how Pakistani educators are navigating the opportunities and challenges of AI in their daily practice.

\section{Method}\label{sec:method}

This study utilized a descriptive survey design to gather data on AI adoption, use, and impact among school educators in Pakistan. A structured questionnaire titled “AI Adoption, Use, and Impact in Pakistan” was developed (using Google Forms) and distributed online to in-service teachers across various regions. Participation was voluntary and responses were collected anonymously in July 2025. The survey instrument was divided into four sections aligning with the study objectives: 

\begin{enumerate}
    \item Demographics, capturing basic information such as gender, school type (public or private), and institution size (number of students/faculty);
    \item AI Adoption, focusing on educator's familiarity with AI tools and their attitude towards adopting AI (including willingness to use AI if support is provided, likelihood of trying new AI tools in the next 6 months, and any formal training received);
    \item AI Use, examining current usage patterns (frequency of using AI in teaching routines, how many days AI was used in the past month, and in which scenarios or tasks AI is employed);
    \item AI Impact, capturing perceptions of the outcomes of AI use (through Likert-scale statements about student engagement, student performance, teacher workload, teaching strategies, and equity of access).
\end{enumerate}
The questionnaire items included multiple-choice and Likert-scale formats, and were designed for an undergraduate reading level to ensure clarity.

A total of 125 educators responded to the survey. Table \ref{tab:KPI-summary} summarizes key performance indicators derived from the survey data (see Section \ref{sec:results}). These respondents represented a range of school contexts: approximately 82\% (n=102) teach in public-sector schools while 15\% (n=19) are from private schools (the remainder did not specify). The sample skewed slightly male (about 60\% male, 39\% female, and 1\% preferred not to state gender). Institution sizes varied; however, over half of the teachers (roughly 61\%) reported working at large schools with 500+ students, whereas only about 7\% came from very small schools with under 50 students. This distribution suggests that urban or well-established schools (which tend to be larger and public) were more heavily represented among respondents. Data from the survey were exported and analyzed quantitatively. We computed frequency distributions and percentages for categorical responses, and calculated mean Likert scores for attitude and perception items. In the Results and Discussion section \ref{sec:results}, we interpret these findings in depth, using simple percentages and mean values to highlight patterns. All quantitative results are reported as proportions of respondents or average scores on a 5-point scale, as appropriate. No identifying information was collected, and the study adhered to ethical standards of informed consent and confidentiality.

\begin{table}[ht]
\centering
\caption{Key Survey Indicators (KPI) of AI Adoption and Use in the Sample (N = 125)}
\label{tab:KPI-summary}
\begin{tabular}{p{10cm}c}
\toprule
\textbf{Metric} & \textbf{Value} \\
\midrule
Active AI Users (uses AI often or daily in teaching routine) & 30\% \\
High AI Familiarity (regularly or daily uses AI in classroom) & 26\% \\
Frequent Lesson Plan Integration (often or fully integrates AI into lesson planning) & 24\% \\
\midrule
Average Agreement with Positive AI Impact Statements\textsuperscript{*} & 3.7 / 5.0 \\
\bottomrule
\end{tabular}
\begin{tablenotes}
\footnotesize
\item *Measured on a 5-point Likert scale (1 = Strongly Disagree, 5 = Strongly Agree).
\end{tablenotes}
\end{table}

\section{Results and Discussion}\label{sec:results}
The analysis in this section is organized around the demographic profile of participants, teacher's attitudes toward AI, their levels of familiarity and usage, the frequency of application-specific use, and the institutional differences in perceptions. 

\subsection*{Participant Profile}
\begin{figure}[h]
  \centering
  \begin{minipage}{0.48\textwidth}
    \centering
    \includegraphics[height=5cm,keepaspectratio]{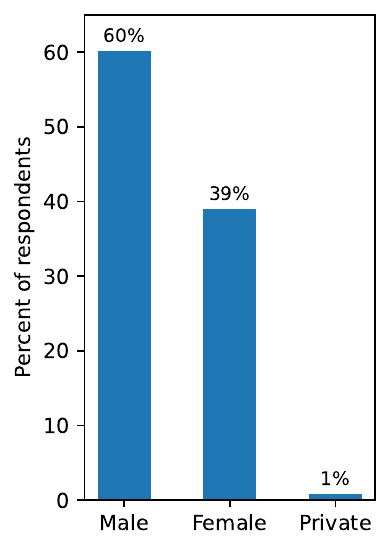}
  \end{minipage} \hspace{-0.22\textwidth}
  \begin{minipage}{0.48\textwidth}
    \centering
    \includegraphics[height=5cm,keepaspectratio]{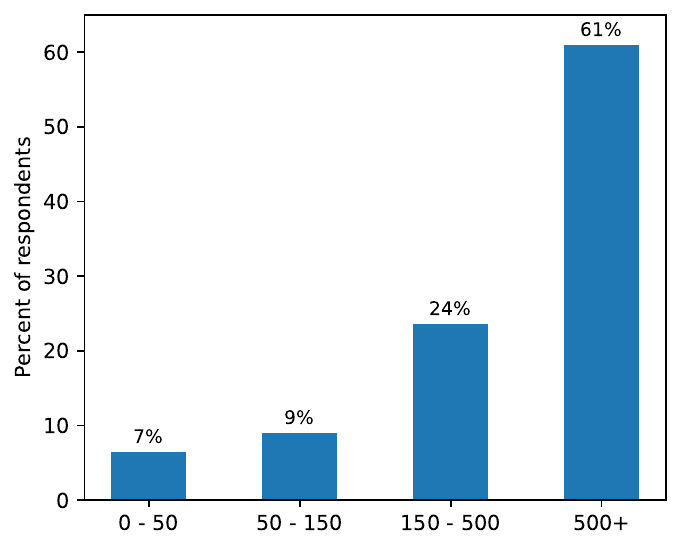}
  \end{minipage}
  \caption{Distribution of respondents by gender (left) and by institution size (right).}
  \label{fig:demography}
\end{figure}

The survey sample consisted of 125 educators representing diverse demographics. As shown in Figure~\ref{fig:demography}, the gender distribution was somewhat skewed, with approximately 60\% male, 39\% female, and 1\% preferring not to disclose. This indicates that while both genders are represented, male educators dominate the sample. In terms of institution size, over 61\% of respondents came from large schools with more than 500 students, while only 7\% worked in schools with fewer than 50 students. The dominance of large institutions suggests that the findings primarily reflect teachers with access to relatively structured and resource-rich environments. These demographic trends are important because exposure to technology often correlates with institutional size and resources, potentially shaping AI adoption behaviors.

\subsection*{Attitudes Toward AI}
\begin{figure}[h]
    \centering
    \includegraphics[width=0.85\textwidth]{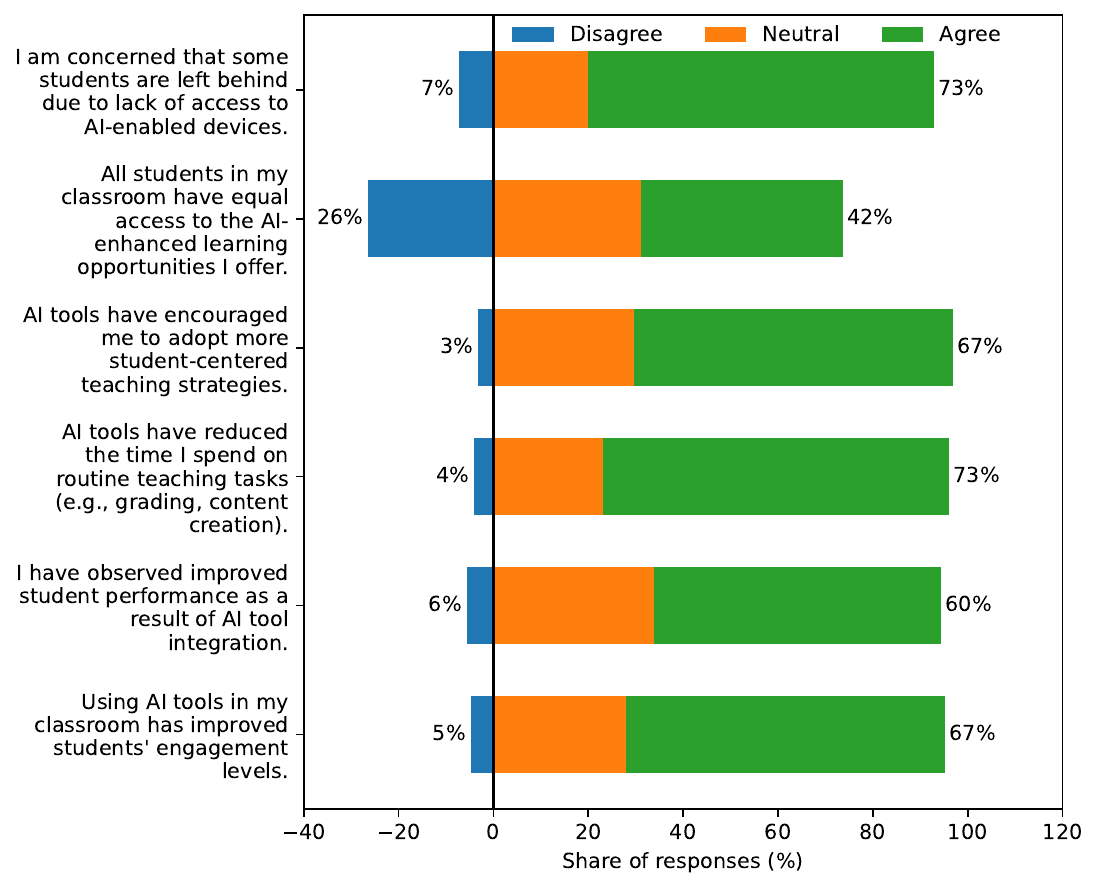}
    \caption{Teacher's perception of AI tools in the classroom, showing the distribution of responses (disagree, neutral, agree) across statements related to access, teaching practices, workload reduction, student performance, and engagement.}
    \label{fig:funnel}
\end{figure}

Teachers expressed broadly positive attitudes toward AI integration. Figure~\ref{fig:funnel} shows the Likert responses to statements regarding engagement, performance, workload, teaching practices, and equity. Approximately 67\% agreed that AI tools improved student engagement, and 60\% observed better performance. Notably, 73\% reported reduced workload through AI use, emphasizing the technology’s time-saving potential. However, equity remains a concern: only 42\% believed all students had equal access, while 73\% feared some were left behind due to lack of devices. These results highlight a dual narrative: AI can enhance teaching efficiency and engagement but also risks deepening inequalities if access barriers persist.

\subsection*{Familiarity and Usage Patterns}
\begin{figure}[h]
    \centering
    \includegraphics[width=0.75\textwidth]{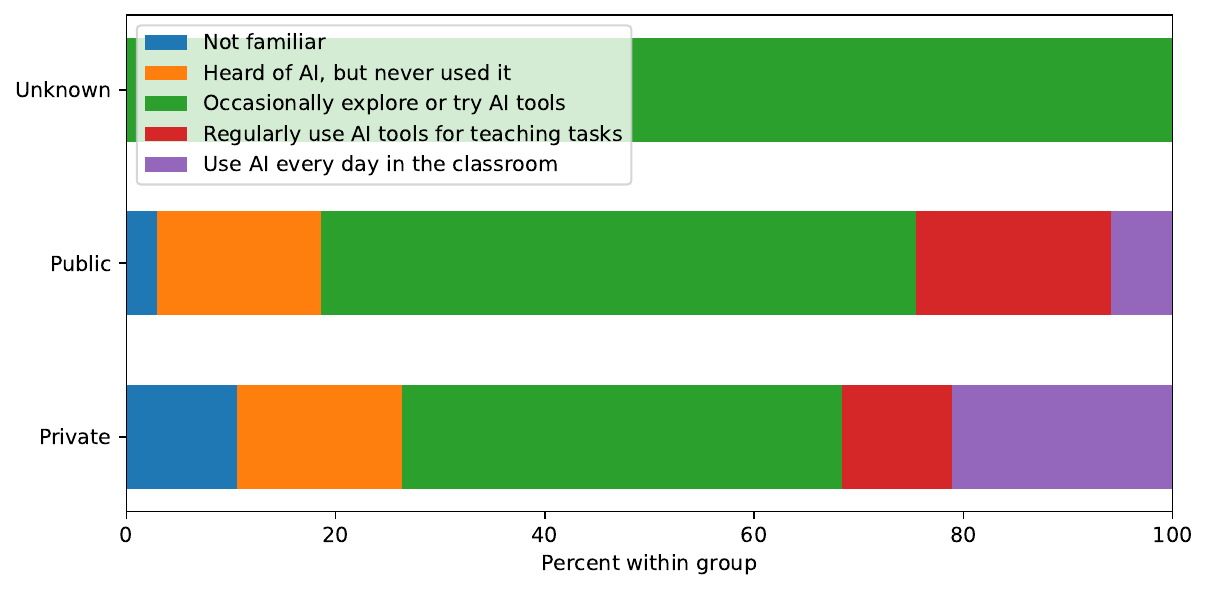}
    \caption{Levels of AI tool adoption among educators across institution types, showing familiarity and usage patterns ranging from not familiar to daily classroom use.}
    \label{fig:ai_adoption_levels}
\end{figure}

Figure~\ref{fig:ai_adoption_levels} illustrates AI adoption across institution types. Public school teachers were more likely to experiment occasionally, while private school teachers showed polarization: some reported daily use while others had no exposure at all. Overall, about 26\% of all educators can be classified as highly familiar with AI (regular or daily use). This suggests that institutional context significantly shapes AI familiarity, with private schools fostering both early adopters and laggards, whereas public schools lean toward moderate, exploratory usage.

\subsection*{Frequency of Application Use}
\begin{figure}[h]
    \centering
    \includegraphics[width=0.55\textwidth]{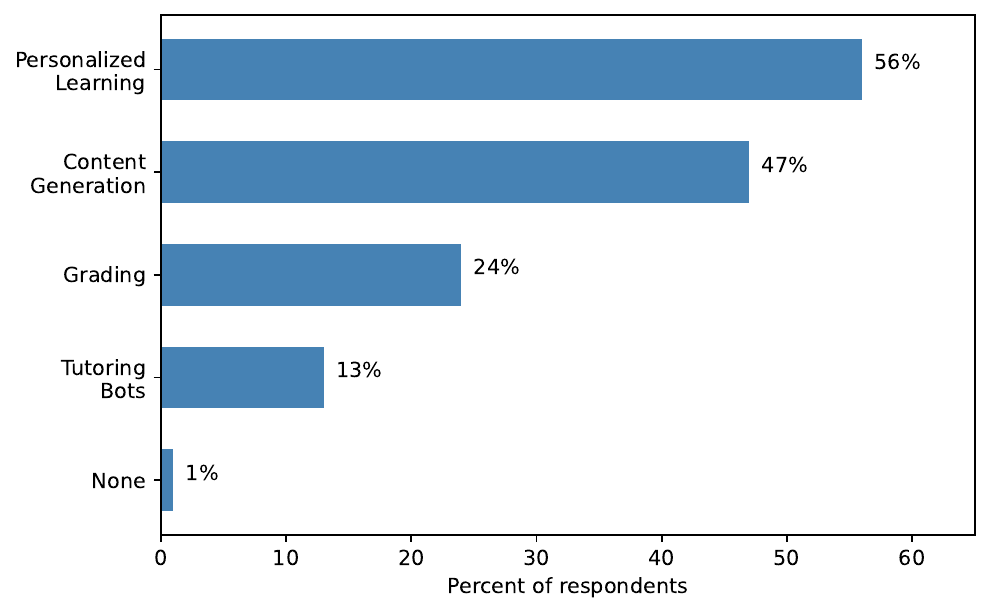}
    \caption{Applications of AI tools reported by educators, with the highest use in personalized learning, followed by content generation, grading, and tutoring bots.}
    \label{fig:likert}
\end{figure}

Educators reported using AI most frequently for content-related tasks. As shown in Figure~\ref{fig:likert}, personalized learning applications topped the list (56\%), followed by content generation (47\%), grading (24\%), and tutoring bots (13\%). This indicates that teachers primarily see AI as a tool to support preparation and customization of learning materials, rather than for evaluative or interactive tasks like grading or tutoring. The hesitance in adopting AI for assessment reflects concerns over fairness, accuracy, and the need for teacher oversight in evaluative processes.

\subsection*{Institutional Differences in Perceptions}
\begin{figure}[h]
    \centering
    \includegraphics[width=0.6\textwidth,trim=0 0 0 60mm,clip]{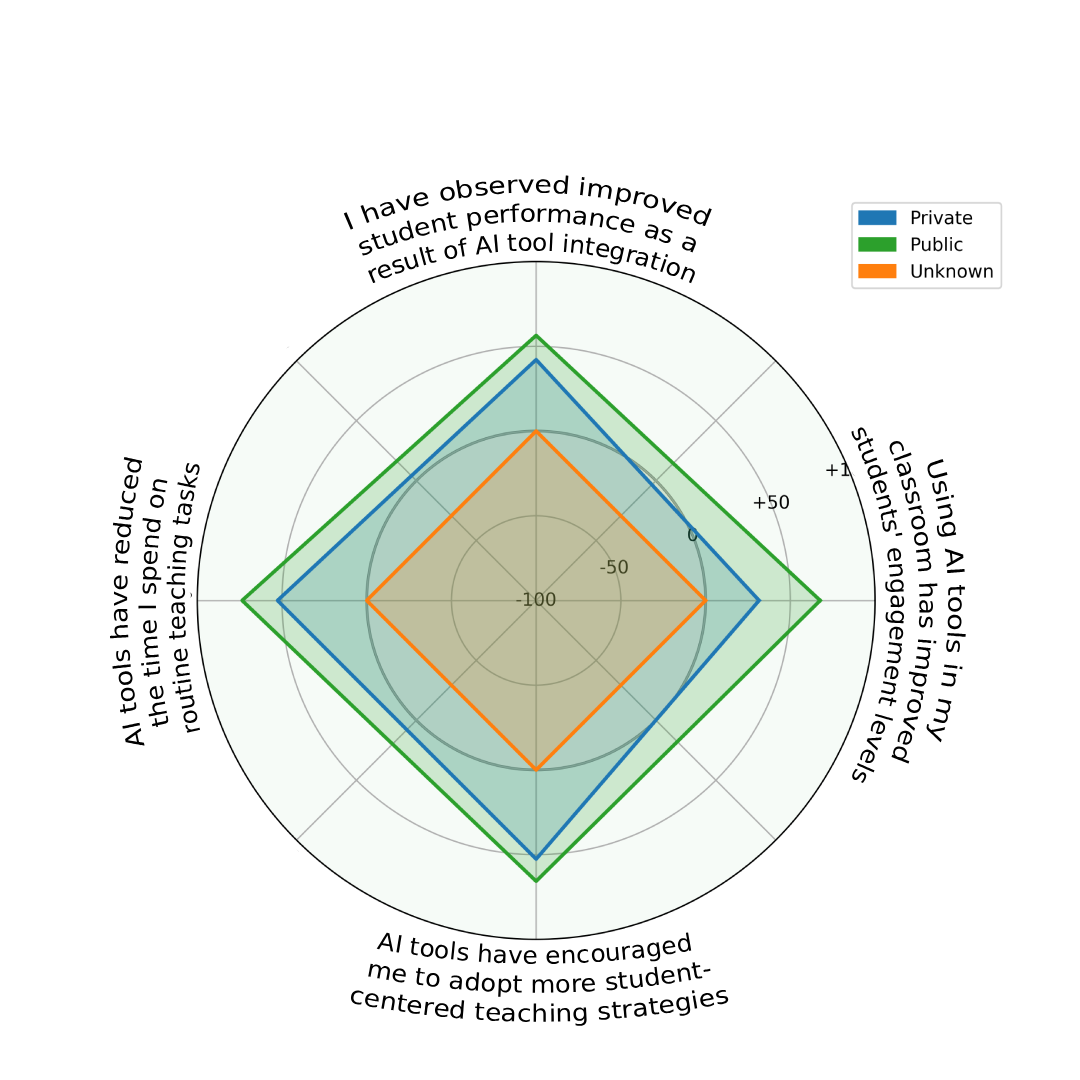}
    \caption{Comparison of teacher's perceptions of AI tool integration by institution type (private, public, and unknown), showing reported impacts on performance, engagement, teaching strategies, and workload.}
    \label{fig:webplot}
\end{figure}

Figure~\ref{fig:webplot} compares teacher perceptions across public and private institutions. Private school teachers reported higher perceived improvements in student engagement and teaching strategies, while public school teachers emphasized AI’s workload reduction benefits. These contrasts reflect differing institutional priorities: private schools may leverage AI for innovation and competitiveness, while public schools focus more on efficiency gains in resource-constrained contexts. Both sectors, however, recognize AI’s transformative potential, underscoring the need for systemic support to harmonize practices.

Taken together, these results suggest that while AI adoption in Pakistan’s schools is still in early stages, teachers recognize its benefits, particularly for content creation and efficiency. Adoption is uneven across institution types, and equity concerns remain pressing. To move from occasional experimentation to deeper integration, policymakers must address access disparities and provide training that aligns with the diverse contexts of public and private schools.

\section{Conclusion}\label{sec:conclusion}

In this study, we examined how K–12 educators in Pakistan are beginning to adopt AI tools, how they are using them, and what impact these technologies are having in the classroom. The findings present a story of cautious optimism. On one hand, Pakistani teachers are clearly interested in and open to AI: a large majority are willing to embrace AI in teaching, especially if adequate support and training are provided. Many have already started using AI, primarily to aid in lesson planning and content creation, and they report valuable benefits such as time savings and more student-centered teaching practices. These early adopters have seen their students engage more and even perform better with the help of AI-driven resources. On the other hand, the integration of AI into everyday teaching remains at an intermediate stage – most teachers are experimenting occasionally rather than using AI in every lesson – and significant challenges temper the enthusiasm. Chief among these is the issue of equitable access: teachers worry that without addressing resource gaps, AI might help some students while leaving others behind, thereby widening existing inequalities. Additionally, many educators lack formal training in using AI tools and thus may not be fully confident or aware of how to harness AI’s potential beyond basic uses.
\section{Limitations and Future Work}\label{sec:limitations}

While this research offers valuable insights into AI adoption in Pakistani schools, it is not without limitations. First, the sample of 125 educators, though diverse, may not be fully representative of all teachers nationwide. The respondent pool was skewed toward large public schools and included a higher proportion of tech-interested educators (given the online voluntary survey format). This means the findings could overestimate AI usage relative to the general teacher population, as those with no interest or access to technology might be under-represented. Additionally, the data are self-reported – teacher's assessments of their AI usage and its impacts might carry subjective bias or optimism. We did not have an objective measure of student outcomes or classroom observations to triangulate the reported impacts on engagement and performance. There is also the possibility of social desirability bias; some respondents might have overstated positive attitudes towards AI knowing it is a “modern” trend. Another limitation is the cross-sectional nature of the survey: it captures a snapshot in mid-2025, which is a time of rapid change in AI tools. Perceptions and usage patterns might evolve quickly as new AI applications (e.g., more advanced tutoring systems or grading tools) become available. Lastly, certain survey items could have been interpreted differently by respondents (for example, what qualifies as “often” using AI might vary between individuals), introducing some ambiguity into the frequency data. These caveats suggest caution in generalizing the exact percentages beyond the study sample, though the overall trends are likely indicative of broader phenomena.

Building on this work, future research should delve deeper into AI’s role in education within Pakistan and similar contexts. A natural next step is to conduct qualitative studies – such as interviews or classroom observations – to explore how teachers are integrating AI and what specific challenges or successes they encounter in practice. Such qualitative insights would complement the broad survey trends and help explain the “why” behind certain patterns (for instance, why some teachers remain neutral on AI’s impact, or how exactly teachers manage when students lack access). Longitudinal studies would also be valuable: following a cohort of teachers over time as they receive training or as new AI tools are introduced could reveal how adoption matures from occasional use to deeper integration. Comparative studies across different regions or school types (urban vs rural, public vs private) within Pakistan could uncover contextual factors that influence AI uptake. It would also be instructive to examine student perspectives and outcomes directly – for example, do students in AI-enhanced classrooms show improved learning gains or motivation compared to those in traditional settings? And importantly, does AI use help bridge or inadvertently widen achievement gaps? Another promising avenue is to evaluate specific interventions: for instance, piloting a teacher training program on AI pedagogies and measuring its effect on teachers’ confidence and usage rates. Given the concern about equity, research on low-cost or offline AI solutions tailored for under-resourced schools would be highly pertinent (e.g., AI tools that can run on mobile phones without constant internet). Finally, as AI in education is a fast-evolving field, continuous monitoring of emerging technologies (like newer generative AI models or adaptive learning platforms) and their adoption by teachers will be necessary. By pursuing these lines of inquiry, future work can support a more evidence-based and inclusive integration of AI into education – helping to turn the current cautious optimism into realized gains for teachers and students alike.

\bibliographystyle{unsrt}  
\bibliography{main}

\end{document}